\documentclass{aa}
\usepackage{graphicx}
\begin{document}

  \titlerunning{XMM-Newton observation of SN1993J in M81}

   \title{XMM-Newton observation of SN1993J in M81}

   \author{H.-U. Zimmermann, B. Aschenbach
          }

   \offprints{H.-U. Zimmermann}

   \institute{Max-Planck-Institut f\"ur Extraterrestrische Physik, 
              Postfach 1312, D-85471 Garching, Germany\\
              email: zim@mpe.mpg.de
             }

   \date{Received  ; accepted  ; draft version 11-Apr-2003}

   \abstract{
In April 2001 SN1993J was observed with both the PN and 
MOS cameras of the {\sl XMM-Newton} observatory, resulting in 
about $7\times 10^4\, s$ of acceptable observation time. 
Fit results with both the PN and MOS2 camera spectra 
studying different spectral models are presented. The 
spectra are best fitted in the energy range between 0.3 
and $\sim$ 10 keV by a 2-component thermal model 
with temperatures of
$kT_1$ = 0.34$\pm$0.04 keV and $kT_2$ = 6.54$\pm$4 keV,
adopting ionization equilibrium. A fit with a shock model also
provides acceptable results. Combining the {\sl XMM-Newton} data 
with former X-ray observations of the supernova, we discuss 
the general trend of $L_{\rm x}\propto t^{-0.30}$ and 
the bump of the X-ray light curve as 
well as former and recent spectral results in the light of 
the standard SN model as first proposed by Chevalier in 1982.
}

   \maketitle


\section{Introduction}

M81 at the moderate distance of 3.6 Mpc 
\cite{Freedman94} is one of the best observed galaxies. When end
of March 1993 the supernova SN1993J exploded in that galaxy, 
there was good hope to learn about the progenitor and
follow, as with SN1987A in the LMC, the evolution of the early 
phases for  
years with high quality spectra in many wave bands.
5 days after the detection in the optical bands (\cite{Ripero93}),
radio emission
from SN 1993J was observed at 22.4 GHz \cite{Weiler93} by the 
VLA, demonstrating the interaction with circumstellar 
material (CMS), an important prerequisite for X-ray emission. 
1 day later, at day 6, the SN was observed with the ROSAT 
satellite at soft X-ray energies around 1 keV (12 Angstrom). 
The immediate detection of a strong X-ray signal (Zimmermann et al.1993a, b)
at these early times was unprecedented up to that date, 
even though in the following years a few of the 18 presently known 
X-ray supernovae\footnote{http://www.astro.psu.edu/$\sim$immler/supernovae\_list.html} 
have been detected at even earlier epochs.
The X-ray satellite ASCA with an 
extended spectral coverage \cite{Tanaka93} and the hard X-ray 
instrument OSSE onboard GRO \cite{Leising94} extended a few 
days later the spectral bands in which emission of SN1993J was 
observed. 

X-ray observations of the supernova were continued with the 
ROSAT observatory up to 1998 and with the ASCA satellite up 
to 1996. After a break of 2 years the CHANDRA observatory 
observed SN1993J in May 2000 \cite{Swartz03} followed by 
the {\sl XMM-Newton} observation in April 2001, reported here.

\section{The Supernova Scenario}
The fact that 93J began as a type II SN, characterized by 
prominent H lines in the optical spectra and few months later 
changed to the Ib type, characterized by no or few hydrogen but 
rich in He, is interpreted that the progenitor had lost all but 
a small amount of its hydrogen layers, thus that the shock heated 
effective photosphere could quickly sink through the thin H layer 
into the deeper He layers during the initial expansion and cooling
phase.

In the case of SN1993J, optical plate material taken before the 
explosion, the optical light curve and theoretical considerations 
indicated, that the progenitor was probably a K0 I red supergiant
with an initial main sequence mass of about $15 M_\odot$ , that had 
kept less than $1\,M_\odot$ of H at the time of the explosion 
\cite{Nomoto93},\ 
\cite{Podsiadlowsky93,Ray93},\ \cite{Bartunov94},\ \cite{Utrobin94},
\cite{Woosley94}, \cite{Aldering94}.

Loss of the outer hydrogen layers of the star may have been 
caused by either a strong stellar wind or by mass (Roche lobe) 
overflow to a companion star in a binary system.
It is known that red supergiants blow slow winds, in the order of 
10 km/s and carrying between $10^{-6}$ to $10^{-4}\ M_\odot$ per year, 
into space. It is unclear whether a wind alone can account for the 
extreme loss of hydrogen in a star of $15\ M_\odot$, therefore 
there is some tendency to prefer a binary scenario for SN1993J with 
mass overflow to a companion, while additional material
escaped as a wind. In a binary scenario one might expect disk-like
asymmetries, which indeed have been found in the profiles of
some optical lines \cite{Matheson00}. On the other hand radio images
of the SN1993J explosion nebula \cite{Bietenholz01} show nearly 
perfect circularity with no indications for asymetries ($< 6 \%$).
 
In the collaps of the inner core of a supergiant star
an extreme shockwave traverses the layers of the supergiant,  
driving the upper layers of the star into a circumstellar material 
producing strong shocks in that medium.
Observations of the shock interaction in the radio, optical and
X-ray bands provide inputs for understanding the explosion dynamics
as well as element composition and mixing. Its physical basis is
usually described in the standard or 2-shock model first worked 
out by \cite{Chevalier82}.

In this model the expanding supernova drives a forward shock with
a typical velocity of $10^4$ km/s into the circumstellar matter
giving rise to temperatures in the order of $10^9$ K, $\sim 100$
keV. The circumstellar matter is assumed to originate from the
stellar wind of the supergiant progenitor star. The resultant
density profile is then a function of the mass loss rate by the wind, 
the wind velocity and has a radial dependance of $r^{-s}$, where s=2 
if the wind is isotropic and constant.
In the model also a reverse shock is forming by the interaction of
the shocked material with the SN ejecta, from which temperatures in
the order of 1 keV are expected. Initially the reverse shock region
may be radiative and can form cool material that may absorb the
X-radiation of that region. The density profile of the SN ejecta is
also described by an $r^{-n}$ dependance. One can then derive a self
similarity solution for that scenario where n and s, describing the
density profiles of the ejecta and the circumstellar medium, play an
important role.

The reason that early X-rays are so rarely seen, up to now only from
18 events, is mainly caused by the circumstellar environment. If the
density is very high, then the initial radiation may not be able to
ionize all the circumstellar material and X-rays may be fully absorbed.
If, on the other hand, the circumstellar matter density near to the
SN is very small then the intensity of the X-rays from the shocks
may be too weak to be detected with present day instruments, as was 
the case with SN 1987A.

\section{{\sl\bf XMM-Newton} observations}
{\sl XMM-Newton} was pointed towards the center of M81 - SN1993J 
is at 3 arcmin distance - on April 22/23, 2001, for a total 
of 132 ksec, from that data of 90 ks with the PN camera in small 
window mode and 83 ks with the MOS2 camera in image mode were
taken. The medium filter was applied both in the PN and the 
MOS2 instrument setup during the observation. 
The MOS1 camera was set to timing mode and not used in this analysis.

Due to high and variable particle background in parts of the 
observing period, a temporal screening had to be applied.
While for the MOS2 camera only the last few 1000 s were screened
off, leaving 77 ksec of good data, in the PN observation about 
half of the observation was affected by bursts in the particle 
background.
In order to maximize the signal to noise ratio in the PN spectrum,
we determined the rates in intervals of about 1000 s separately for 
the source plus background field and the background field and 
eventually excluded all time intervals that deteriorated the
signal to noise ratio in that band, leaving 69 ks of good data. 
The optimization relative to the cut radius of the source plus 
background field and the deselection of background affected time
intervals was performed in different energy bands. The final 
selection of accepted intervals was done in the band above 
1 keV to optimize the signal in the part of the spectrum
where the lines clearly stand out in the spectrum. The cut radius 
chosen was 300 pixels, corresponding to 15 arcsec. The spectrum was 
binned such that each bin contained at least a fixed number of 5
(background subtracted) source counts.

\section{The {\sl\bf XMM-Newton} spectrum}

The PN spectrum of SN1993J shows emission lines of highly ionized 
Mg, Si, S, Ar, Ca and very cleary the complex of the Fe lines at high 
energies.
Spectral fits were first performed using only the PN camera data
offering the best statistics.
Fitting was tested with different model components. Good fits were 
achieved over the whole energy band between 0.3 up to $\sim$ 11 keV,
using a 2-component thermal model with variable 
element abundances (vmekal in the nomenclature of XSPEC). We also 
tried different shock 
models and found that a sedov model with variable abundances 
(vsedov in XSPEC) produces also acceptable $\chi^2$ 
values, although all tested shock models 
tend to reproduce not so well the 
different line complexes visible in the spectra above 1 keV.

\begin{figure}
   \includegraphics[angle=-90,width=8.8cm,clip=]{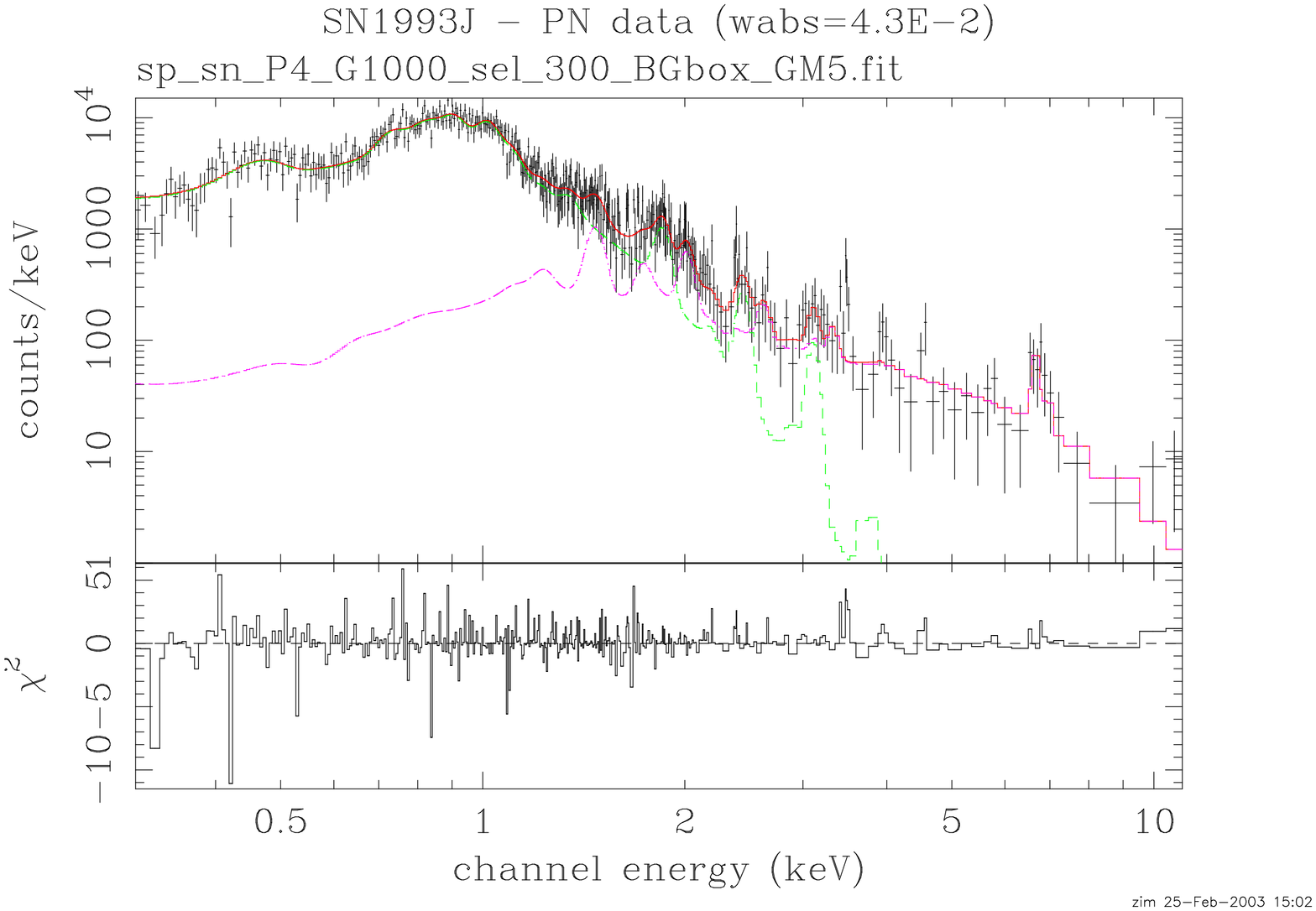}
    \caption[]{{\sl XMM-Newton} spectrum as observed with the PN camera and 
           the best fit thermal 2-component model with $kT_1$=0.34 keV 
              and $kT_2$=6.54 keV.
           The dashed curves show the low and high
           temperature component.
              }
       \label{sp_PN}
  \end{figure}

An overview on the best fit parameters with both the PN and the
MOS2 camera and for different models is given in Tab. 
~\ref{fit_temperatures}.
The PN camera spectrum and the best fit 2-component thermal model
(2vmekal) is shown in Fig. ~\ref{sp_PN}.

The spectral differences found between the PN and MOS2 fits can be understood 
as uncertainties in the spectral cross calibration of the two instruments
(as stated in the document 'XMM-EPIC status of calibration and data analysis',
XMM-SOC-CAL-TN-0018 2003, valid for the software version SAS 5.4 applied 
in this analysis).
In the following we therefore take the best fit results from 
the PN camera with its better statistics (7515 counts compared 
to 2593 from MOS2), especially at higher energies more 
representative for the overall behaviour, as the basis for the 
spectral discussion. 

\begin{table*}
      \caption[]{Selected best fit parameters of different models.
  2vmekal is a 2-component thermal model with variable element abundances,
  2vmekal1 is the same model but with identical element abundances for the
  2 temperature components, and vsedov is a shock model with 
  variable abundances.
        }
         \label{fit_temperatures}
{\small

\begin{tabular}{l|lllllll}
instrument\ \ \  &model\ \ \ \ & $kT_1[keV]$\ \ \ &$kT_2[keV]$\ \ \ & 
          $N_{H1}[10^{22}cm^{-2}]$ & $N_{H2}[10^{22}cm^{-2}]$ & dof\ \ \  & $\chi^2_r$\ \ \ \\
\hline
PN      &2vmekal & $0.34^{+0.05}_{-0.03}$ & $6.54^{+4}_{-4}$      & 0.55$\pm0.21$ & 0.33$\pm1.80$ & 360 & 0.90 \\ 
PN      &2vmekal1& $0.35^{+0.04}_{-0.04}$ & $4.24^{+3.8}_{-1.8}$  & 0.45$\pm0.13$ & 0.18$\pm0.75$& 371 & 0.94 \\ 
PN      &vsedov  & $ 0.51\pm0.04$          & $7.0\pm0.2$        & 0.17$\pm0.037$     &              & 376 & 1.01 \\
MOS2    &2vmekal & $0.59^{+0.03}_{-0.04}$ & $3.48^{+1.2}_{-1.2}$  & 0.25$\pm0.25$ & 0.0$\pm0.54$ & 113 & 1.25 \\
PN+MOS2 &2vmekal & $0.33^{+0.46}_{-0.46}$ & $4.09^{+1.2}_{-1.2}$& 0.59$\pm0.25$   & 0.2$\pm0.8$  & 505 & 0.99 \\ 
\hline 
\end{tabular}
              }
\end{table*}

The Fe complex near to 6.7 keV is represented in the PN camera 
with about 33 counts (in the MOS camera about 5 counts).
The width of the Fe complex in the PN spectrum (extending from 
6.5 to 7.2 keV) is generally better
fit with a higher 'high temperature' component, while other line
complexes, or better the regions between them tend to be better 
fit by somewhat lower temperatures.
But the statistics in the spectrum do not reasonably suggest
to introduce more components with additional temperatures. 

Let us first look at the results from the 2-component thermal 
model fit (2vmekal).
The low temperature component of $kT_1 \sim 0.34 \pm 0.04$ keV 
dominates the spectrum up to energies of about 2 keV, 
but a higher temperature component is definitely needed 
to explain the spectrum at higher energies.

Given the fact that the high temperature component takes into 
account just the band between 3 keV and $\sim$8 keV, where
the spectral fit anchors on the Ca IXX, XX and the prominent
Fe XXIII, XXIV, XXV line complex, the temperature is likely 
to be not very well defined.
Indeed the temperature of the high temperature component of
$kT_2\ \sim 6.4 \pm{4}$ keV is poorly restricted, 
because every change in the temperature component is  
compensated by changes in the abundances of Fe, Ca and Ar. 
Due to this dependence all the element abundances in the 
fits have errors typically much larger than the parameter values
themselves.Therefore, the abundance values should be considered 
with some care.  

\begin{table}
      \caption[]{Best fit element abundance values  
          from the fit to the PN camera data
        }
         \label{fit_abundances}
{\tiny
\begin{tabular}{l|lllll}
el. & 2vmekal\_T1 & 2vmekal\_T2 & 2vmekal1 & MOS2\_T1 & vsedov \\
\hline
He &  15$\pm95$    &   0.1$\pm273$ &  0.3$\pm12$  &  0.0$\pm156$& 259$\pm4881$\\
C  &  0.0$\pm172$  &    -          &  0.0$\pm0.1$ &  0.0$\pm156$&  0.0$\pm$\\
N  &  0.0$\pm172$  &    -          & 140$\pm553$  & 212$\pm1908$&  6.5$\pm120$ \\
O  &  0.15$\pm1.1$ &    -          &  0.2$\pm0.9$ &  0.0$\pm10$ &  1.1$\pm20$\\
Ne &  2.0$\pm11$   &  0.0$\pm$     &  2.5$\pm10$  &  0.0$\pm14$ &  4.0$\pm75$ \\
Mg &  0.4$\pm2.6$  &  47$\pm4531$  &  0.2$\pm1.0$ &  2.0$\pm16$ &  6.4$\pm119$ \\
Al &  0.0$\pm$     &  103$\pm10000$&  0.0$\pm   $ & 13$\pm104$  &  -          \\
Si &  3.6$\pm20$   &  20$\pm1930$  &  4.7$\pm20$  & 3.7$\pm30$  &  7.0$\pm131$ \\
S  &  15$\pm85$    &  10$\pm968$   &  8.4$\pm35$  & 8.2$\pm68$  &  0.5$\pm10$\\
Ar &  132$\pm736$  &  18$\pm1777$  &  16$\pm68$   & 0.6$\pm43$  &  5.8$\pm   $\\
Ca &  95$\pm541$   &  3.7$\pm357$  &  11$\pm46$   & 18$\pm169$  &  0.0$\pm21$ \\
Fe &  1.1$\pm5.9$  &  1.3$\pm121$  &  1.0$\pm4.0$ & 1.8$\pm15$  &  4.8$\pm90$ \\
\hline 
\end{tabular} 
}
\end{table}

One major difficulty with fitting the vsedov model is 
based in its enormous demands on computational power. It is
therefore difficult to properly scan the parameter space.
In the vsedov model the best fit temperatures are 
somewhat higher than those in the vmekal models. 
The fit requires additional intrinsic
absorption.
The lines at higher energies are not very well reproduced by the model.

\section{Discussion}

How do the X-ray observations fit into the scenario for 
SN1993J ?

\subsection{Earlier X-ray spectra}

We have reevaluated the ROSAT data in a homogeneous way
\cite{Zimmermann97} and fitted the spectra with a 
thermal 1-component model (vmekal in XSPEC) where the element 
abundances were fixed to either solar or to the values 
obtained from a similar 
fit to the {\sl XMM-Newton} PN spectrum. Due to the high 
temperatures in the early ROSAT observations there is almost 
no difference between fitting either solar or {\sl XMM-Newton}
 based element abundances. 
In the November 1993 measurements the fit with the thermal 
model and solar abundances turned out to be poor but improves significantly 
(see Tab. ~\ref{ROSAT_temperatures})
with the {\sl XMM-Newton} based element abundances. 

The first ROSAT PSPC spectra do only allow 
to set a lower limit of $>17$ keV for the temperature of 
a thermal bremsstrahlung, Raymond \& Smith or mekal  
model. From the ASCA measurements
a lower limit of $\ >\,10$ keV was reported. 
Around the same time the hard X-ray instrument OSSE
on the GRO satellite could determine a 
temperature of $82_{-29}^{+61}$ keV at day 10 to 15 and, 
statistically already very weak, of $74^{+120}_{-40}$ keV 
about 1 month after the event \cite{Leising94}. 
ROSAT measurements half a year later (Zimmermann et al. 1993d, 
1994a) revealed a strong decrease from the initial high
temperatures to temperatures around 1 keV. An ASCA measurement 
\cite{Uno02} confirmed that strong temperature drop. 

In terms  of the standard
2-shock model it is suggested that 
the observed emission initially originated from the fast 
forward shock, while emission from the reverse shock 
region was blocked at that time by absorbing material due 
to fast cooling 
processes in the denser environment. It has been assumed
that the initial absorption  
disappeared on a time scale of the order of 100 days so that 
half a year later the measured flux was then dominated by the 
emission from the reverse shock region \cite{Zimmermann94, 
Zimmermann96, Fransson96}.

\begin{table}
      \caption[]{ROSAT spectra: best fit parameters of the fit to a 
     1-component thermal model (vmekal in XSPEC). Abundances were 
      fixed to either solar or to those that resulted from a fit 
      to the {\sl XMM-Newton} PN spectrum using the same model.
        }
         \label{ROSAT_temperatures}
{\small

\begin{tabular}{l|lllll}
date & kT [keV] & $N_H [10^{22}cm^{-2}]$ &dof& $\chi2_r$ & abund. \\
\hline
April 93    & $>$ 17         & $0.073^{+0.015}_{-0.012}$ &6& 0.97   & solar\\ 
April 93    & $>$ 18         & $0.071^{+0.005}_{-0.015}$ &6& 0.96   & PN\\ 
Nov. 93 & $1.60^{+0.2}_{-0.2}$ & $0.057^{+0.015}_{-0.015}$ &6& 2.46 & solar\\ 
Nov. 93 & $1.05 ^{+0.16}_{-0.13}$ & $0.14^{+0.03}_{-0.06}$ &6& 0.90 & PN\\ 
April 94 & $0.56^{+0.5}_{-0.3}$ & $0.70^{+0.07}_{-0.03}$ &6& 1.05 & solar \\ 
April 94 & $1.05 ^{+0.16}_{-0.13}$ & $0.14^{+0.03}_{-0.06}$ &6& 0.90 & PN\\ 
\hline 
\end{tabular}
              }
\end{table}

Unfortunately the X-ray lightcurve of SN1993J is not covered between 
days 50 and 200, so that the transition from the forward shock
dominance to the reverse shock remains an open issue.

The Chandra spectrum of May 2000 was fitted by \cite{Swartz03} 
with a 3-component 2vmekal+mekal model. Their best fit model
(kT's  of 0.35, 1.01, and 6.0 keV) does not fit the {\sl XMM-Newton} PN 
data very well, especially in the energy regime below 0.7 keV, 
the low and high temperatures are almost identical.  

\subsection{The X-ray light curve}

  \begin{figure}
   \includegraphics[angle=-90,width=8.8cm,clip=]{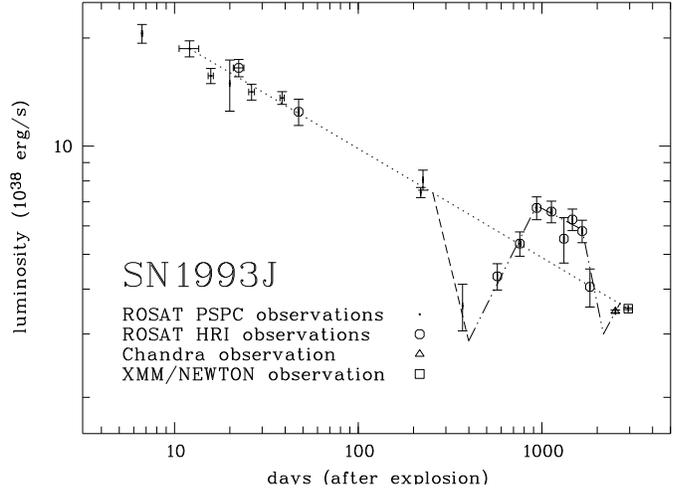}
    \caption[]{X-ray light curve showing the development of the
   X-ray luminosity in the 0.3-2.4 keV energy range. The dotted 
   line indicates a decline rate with $t^{-0.30}$.
   The dashed lines result from model calculations involving 
   time limited decrease and increase of the density in the 
   ambient matter. 
              }
       \label{lightcurve}
  \end{figure}

The lightcurve in Fig. ~\ref{lightcurve} shows the development 
of the X-ray luminosity in the energy range 0.3 - 2.4 keV 
as determined from the 19 ROSAT observations, the Chandra 
observation from May 2000  
and the new {\sl XMM-Newton} data. The tendency
over the first half year is characterized by a decline with
$t^{-0.30}$ (Zimmermann et al. 1993c, d), where t is the time since 
the outburst. Thereafter
the lightcurve shows a bump and 5 years after the outburst the
luminosity appears to return to the initial decline profile. The
CHANDRA observation of 2000 and the {\sl XMM-Newton}  observation in 2001 
are very close to the general $t^{-0.30}$ decline.

The mass loss rate of the wind of the progenitor can be expressed 
as  $\dot{M} = 4\pi\rho v_wr^s$, where $v_w$
is the wind velocity. For a constant and homogeneous wind $s$ has 
a value of 2. Under the assumptions that the density $\rho$ of 
the circumstellar material at radius $r$ is dominated by that wind 
and the shock velocity stays constant, the observed luminosity 
is roughly proportional to the square of the density 
integrated over the emitting volume. In case of a constant and
homogeneous wind the luminosity should show a time dependence of 
$L_x \propto t^{-1}$. The flatter time profile $t^{-0.30}$ of the
ROSAT luminosity observed during the first half year, corresponds for
a spherically symmetric scenario to a density profile with s=1.65.
In a more detailed consideration in the framework of the formalism
of the standard model the development of the
X-ray luminosity produced in the forward shock is expressed by:
$L_x\ \propto\ T_e^{0.16} t^{(3-2s)(n-3)/(n-s)}$ 
\cite{Fransson96}, where $r^{-n}$ describes the radial dependence 
of the density of the SN ejecta. For $n >> 10$ it follows 
$L_x\ \propto\ t^{-(2s-3)}$ and taking the observed slow decline 
with $t^{-0.30}$ the  value for $s\ \approx\ 1.65$ is confirmed,
in good agreement with results published earlier by \cite{Immler01}. 

In the picture of the standard model the shock front, 
while running outward, is probing
increasingly older periods of the progenitor wind with a
velocity roughly 1000 times the wind velocity. Thus in 5 years
ROSAT scanned about 5000 years of wind history.
In our simple formalism the measured time dependence of the 
luminosity requires that the ratio 'mass loss rate/wind velocity' 
decreased during the last 5000 years in the life of the 
progenitor by more than a factor of 7. This scenario has been
investigated in more detail by \cite{vandyk94} using radio data 
and by \cite{Immler01} using the ROSAT data to calculate 
absolute mass loss
rates. The latter authors also speculate that the change of the
wind properties might indicate a transition from a red to a blue
supergiant phase where mass loss rates are at least an order of
magnitude lower and wind velocities may reach  1000 km/s.

The bump in the X-ray light curve suggests a local 
increase in density above the general power-law profile of the 
circumstellar matter. 
It could be attributed to a change in the wind parameters of 
the progenitor or to an asymmetry caused by a possible binary 
scenario. 

Interesting is a correlation with the expansion velocity of the 
SN shell that has been derived from high resolution radio 
images of SN1993J. In Fig. 6
of \cite{Bartel02} 
at about 350 days after the outburst the expansion rate 
of the radio image size shows a break and continues thereafter at
a slower rate. Between day 369 and 1655, the leading edge and 
the plateau of the 
bump, the expansion velocity decreases in the order of 28\%. 
The total mass which gave rise to the observed X-ray luminosity 
bracketing the bump is estimated to   
$1\ M_\odot$. The total mass of the expanding shell 
covered up to day 350 was of the order of $3\ to\ 4\,M_\odot$.
Both features, the radio deceleration break and
the bump in the ROSAT light curve, independently indicate a 
flattening of the density profile in the circumstellar medium. 
Because the break in the expansion curve does not visibly 
affect the circularity of the radio images one may argue that
the cause more likely can be attributed to a density change in 
the wind instead of an asymmetry by a binary scenario that would
possibly affect the circularity of the images.

The closer inspection of the X-ray light curve shows a rapid decline 
between day 225 and day 370, followed by a rapid increase until day 935. 
A similar behaviour could have occured between day 1126 and day 1464, 
but there is just one data point in that period, the upper limit of which 
is consistent with no change. The rapid decline after day 1655 is very 
evident and recovery indicated by the Chandra data point at 
around day 2500 has occured. The existence of the latter two dips 
is statistically not too convincing, but they may be real as we show 
below. We suggest 
that each of the three dips indicates a density depression which is followed 
by a rise of the density. If the shock wave runs into a sufficiently 
low density regime no more matter will be shocked and heated, the 
matter heated so far will simply expand. If the expansion is isothermal  
the light curve will go down with t$\sp{-2}$; if the expansion is 
adiabatic the decline will go like t$\sp{-8/3}$, as long as the temperature 
is higher than about 2 keV. This temperature is about what we  
would expect at around day 350 (cf. Fig. ~\ref{temperature_model}). 
The falling dashed lines in figure 3 represent such expansion. 
After some time the shock wave encounters a jump in density to 
significantly higher values. The density increase raises the emission 
measure and the X-ray luminosity. Initially the luminosity 
increases linearly with t when it hits a step like density jump. 
The rising lines in figure 3 represent such density jumps. 
The intersection between the rising and falling lines denote 
the position or epoch of the density jump. We have compared 
these epochs with those where the highest, discontinuous 
deceleration of the shock wave has occured, which have been derived 
from the expansion rate of the radio images (cf. Fig. 6 of \cite{Bartel02}). 
There are also just three such events in the radio images. 
The two sets of epochs agree surprisingly well with each other to within 
less than 30 days. 
We conclude that these changes in the X-ray light curve reflect 
changes of the density in the ambient matter profile, which appears 
to show some repetitive pattern. Each of these density increases
is preceded by some density decrease like in a wave. 
Whether this characterizes the activity of the progenitor star 
as far as the mass loss is concerned remains to be seen.         

\subsection{Spectral development}

  \begin{figure}
   \includegraphics[angle=-90,width=8.8cm,clip=]{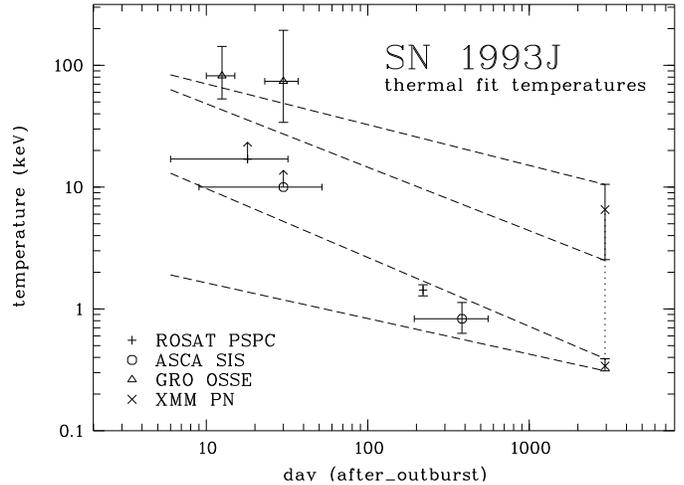}
    \caption[]{X-ray determined emission temperatures. The dashed 
    lines outline the time evolution of the low and high  
     temperature component using the standard model. For each 
     component there are two lines bracketing the range of  
     the temperatures allowed by the {\sl XMM-Newton} data using the 
      2vmekal model.
              }
       \label{temperature_model}
  \end{figure}

We can now combine the {\sl XMM-Newton} temperatures with the ROSAT, ASCA and
GRO results. Let us assume for a moment that the initial
high temperature, as measured by GRO, originated from the forward
shock, while half a year later the flux was dominated by emission from
the reverse shock because of the higher electron densities in 
that region. One can use the similarity solution of the standard 
model to express in general terms the ratio between the temperatures 
in the forward and the reverse shocks that depends only on the 
parameters s and n, 
$T_r/T_f\ =\ (3-s)^2/(n-3)^2$
\cite{Fransson96}.

With the wind density profile of s=1.65 derived from the light curve,
and assuming that the 2 temperatures determined from the thermal 
2-component fit to the {\sl XMM-Newton} spectrum
represent the temperatures in the forward and reverse shock regions,
we obtain n=8.9 for the density profile of the ejecta. Taking into account 
the uncertainties of the temperatures as quoted for the PN 2vmekal model, 
then n results between 6.4 and 10.9. 

Entering s and
n into the formula that describes the temporal behaviour of the shocks,
$T_r\propto\ T_f\ \propto\ t^{-[2(3-s)/(n-s)]}$ \cite{Fransson96}, we can 
extrapolate the {\sl XMM-Newton} temperatures backward in time and compare them 
with the ROSAT, ASCA and GRO results
(see Fig. ~\ref{temperature_model}).  The temperature development
of the forward shock component is roughly consistent with the temperatures 
deduced from the GRO data, and also the low temperature development,
matching the ROSAT data at day 210, is consistent with its interpretation 
as coming from the reverse shock region.
 
But it has to be repeated that, while the parameter s appears reasonably
well determined from the light curve (at least for certain periods),
the value of n is derived under the assumption of thermal equilibrium between
electrons and ions in the emission region. Estimates 
(see \cite{Fransson96}), however, show that even near to the time of the explosion 
equipartition is not expected in the wind shock.  

One can not exclude that we see just
emission from the forward shock. The hot material heated in the
early days has cooled as it expands, both radiatively and
adiabatically with the continuum temperature dropping faster than
the higher ionization levels indicate, i.e. we have a delay of the
recombination because of the long recombination time scales involved.
Thermal equilibrium by just Coulomb collisions 
and ionization equilibrium is no more reached at times later than
about a few days after outburst. Stellar wind matter heated at later 
stages as
the forward shock proceeds is not in thermal equilibrium if not
processes other than Coulomb collisions are at work. Therefore the
electrons would stay relatively cool and also ionization equilibrium
is far from being reached leaving the plasma underionized, which
would explain the observation of the low temperature component.

With the sedov model we have tested such a scenario with emission from 
just one shocked region. But the relatively poor fitting results for the 
line complexes at higher energies show also that this model is not 
quite adequate.

\section{Conclusions}

The {\sl XMM-Newton} spectrum is well fitted by a thermal 2-component 
model with ionization equilibrium. The 
predictions of the standard 
model describe reasonably well the observed 
development of the X-ray spectral temperatures, involving 
both forward and reverse shocks. The equally acceptable fit with a 
vsedov model shows that at this stage one
cannot exclude that only one shock component is at work.

The strong interdependence between the temperature and the values 
for the abundances of the dominating elements, as far as the high 
temperature component is concerned, makes it very difficult to interpret the
abundance results. But there is clearly no need to involve high values for the 
metal abundances, which would point to ejecta heated by the reverse 
shock. 

The new {\sl XMM-Newton} data point in the X-ray lightcurve gives arguments to
the assumption of a general decline rate of $L_x \propto t^{-0.30}$ 
and it suggests that the bump in the light curve 
is probably due to an intermediate density fluctuation. The detailed 
inspection of the X-ray light curve shows that a density increase did not  
occur just once, but in some repetitive fashion, There are 
three dips in the X-ray light curve each one followed by an 
increase in luminosity. Each of the dips occurred very close to the 
times when the forward shock velocity underwent a dramatic deceleration 
as indicated by the expansion rate of the radio images. 
 
The observations show that spatial changes of the powerlaw density
profile have no significant impact on the morphology of the radio images,
that remain very much circular.
This indicates that the bumps in the X-ray lightcurve are more likely caused 
by density changes in a spherically symmetric wind rather than by 
asymmetries introduced by a postulated binary companion. For most of the
time we need a wind that in the past left more material per volume
unit in the circumstellar environment than near to the time of 
explosion.


\end{document}